**Tuning nanosecond switching of spin-orbit torque driven magnetic tunnel junctions**

Shengjie Shi and R.A. Buhrman

*This manuscript was not finalized due to the passing of Prof. Robert Buhrman in April 2021. One year after I decided to upload it to Arxiv. By making it public, I wish that this pioneer work in fine tuning SOT switching in MTJ can benefit researchers in this field, and wish to memorize Bob for his years devoted to spintronics research as well as the people and the lab. We all miss you very much.*

## Abstract

**Since the discovery of the spin orbit torque (SOT) induced by spin Hall effect in heavy metals, much effort has been devoted to understanding the mechanism of the charge-to-spin conversion as well as to developing new schemes for high speed, low energy magnetic recording technologies. While fast switching has been demonstrated in three terminal magnetic tunnel junctions (3T-MTJs) through applying short voltage pulses in the heavy metal channel, detailed understanding of the switching mechanism is lacking due to the complexity of the multi-layered magnetic structure and the three-terminal geometry. We show in this letter that current-induced effective fields play a key role in the fast switching and by tuning the applied external field we can finely tune the symmetry of the pulse switching between two switching polarities, namely parallel to anti-parallel (P-AP) and anti-**

**parallel to parallel (AP-P). These results show that the manipulation of detailed magnetic configuration is the key to fast switching and is a useful way for future optimization of SOT memory and further applications.**

Since the discovery of spin Hall effect and its application in SOT driven 3T-MTJs, there have been continuous efforts in optimizing it to achieve lower switching current (in-plane magnetized MTJs), or field-free switching (perpendicularly magnetized MTJs), or fast, nanosecond switching (both types). Reports have shown that the switching dynamics in SOT driven nano-structures can be more complicated than a single domain, macrospin model description when there is presence of domain structures, Dzyaloshinskii-Moriya interaction (DMI), shape-induced edge magnetization, etc. Previous works have mentioned that effective fields associated with electric current, including the Oersted field and the field-like field, are potentially beneficial to the fast magnetization switching[1–3]. In this letter we study the pulse switching behavior of a large number of 3T-MTJ devices on various spin Hall channels and demonstrate that these effective fields, which are collinear with the free layer anisotropy direction, induce non-trivial spin dynamics and can alter the symmetry of the magnetization switching.

The samples are grown in a Canon Anelva chamber with a stack structure W(4)-Hf(0.25)-FeCoB(1.8-2.1)-Hf(0.1)-MgO-Capping (hereafter called W samples). Numbers in the parentheses are thicknesses in nanometers. The samples are fabricated into 3T-MTJs using the same method mentioned in our previous works[1]. Figure 1a shows a typical W device geometry and the measurement scheme. The device shows clean, abrupt switching of the free layer (FL) with either external field sweep (Fig. 1b) or electric current sweep in the SHE channel (inset of Fig. 1b). The horizontal shift of the field switching minor loop is caused by residual dipole field from the reference layer, which is not fully cancelled by the synthetic antiferromagentic layer. In order to investigate the impact of the effective field on the pulse switching behavior, we tune the applied field around the apparent dipole field so that the dipole field is either not fully compensated or overly compensated. We carry out pulse switching measurements in which we apply short voltage pulses to the SHE channel with amplitude varying from 0.38 V to 0.6 V and duration varying from 0.5 ns to 8.5 ns, and measure switching probabilities at each combination of pulse amplitude and duration. We plot switching probability curves in Fig. 1c-1h for three consecutive applied fields, $H_{appl}$ = -20 Oe, -4 Oe and 15 Oe. We observe a clear trend of shifting symmetry from favoring AP-P at H = -20 Oe to favoring P-AP at H = 15 Oe. We also see that the most significant change lies in the low voltage curves where a change of applied field either raises or lowers the probability dramatically, which signifies the tuned onset of the spin torque switching. Table 1 summarizes the $V_0$'s and $t_0$'s at different applied fields, where $V_0$ and $t_0$ stand for critical switching voltage and characteristic switching time respectively, as acquired from a macrospin fit $V=V_0(1+t_0/t)$. As the field is changed from -20 Oe to 15 Oe, $V_0$ increases for AP-P and decreases for P-AP, which matches well with the symmetry change seen in the switching probability curves. The above results strongly show that we can tune the symmetry of the AP-P and P-AP switching by tuning the field in a direction collinear with anisotropy direction. Since the current induced effective fields either strengthen or diminish the apparent dipole field in the direction of the FL anisotropy in our 3T-MTJ systems, our result is a strong evidence that these fields are critical to the dynamics of the switching.

While tuning the applied field by a large amount (~ 15 Oe) induces symmetry changing of the $V_0$ in the pulse behavior, fine tuning of the applied field within 10 Oe also brings changes of the pulse switching behavior in a more trivial way. Table 2 shows the same device as in the previous paragraph pulsed at $H_{appl}$ = -10 Oe, -4 Oe and 4 Oe. When the change of the external field is small, $V_0$ stays almost unchanged since it does not impose significant increase/decrease in the spin torque to switch the FL. However, characteristic switching time changes monotonically with field, which also indicates change of switching energy due to the unbalanced field. This cannot be explained by the macrospin model since the $t_0$ should be largely dependent on the FL anisotropy $H_k$, which is significantly larger than the small change of the applied field. Micromagnetic simulation has shown that the curling states at the edges of the nanomagnet and initial magnetization configuration add complexity to the short pulse switching behavior[4–7]. Due to our elliptical device geometry, we speculate that a small change of field in the device plane can change the initial excitation of the FL which changes domain nucleation speed. Macrospin model shows that the characteristic switching time is inversely proportional to $\alpha$, $\gamma$ and $H_k$, where $\alpha$ is magnetic damping, $\gamma$ is gyromagnetic ratio and $H_k$ is the anisotropy field of the FL[8]. Using typical values from our MTJ structures, it gives $t_0 \sim 30$ ns, which is more than an order of magnitude larger than our results. This discrepancy is also likely due to the domain structure that causes fast nucleation and propagation different from coherent rotation of the magnetization. The trivial change of the switching behavior can also be confirmed by write error rate measurements. Figure 2 shows the 1ns duration write error rate carried out on the same device at $H_{appl}$ = -10 Oe (blue), -4 Oe (black) and 4 Oe (red). While the horizontal axis is the normalized switching voltage $V/V_0$, which removes the dependence of the WER on the $V_0$, apart from the fast roll-off of each curve which shows excellent reliability, the WER behavior still shows symmetry changing when we vary the applied field. This behavior is not only seen in W samples, but also in many other material systems that we have studied (supplemental material). They together demonstrate that the switching mechanism in 3T-MTJs cannot be solely described using a single magnetization model, and an in-plane effective field alters the dynamics of the switching and is a possible factor for fast, nanosecond switching time scale that we widely see in 3T-MTJ structures. Further micromagnetic modeling and fine device geometry alternation can be the next step to reveal the details of FL switching under short voltage pulses.

To confirm the existence of the current induced effective fields in our devices, we run bias current in the channel and measure a series of field switching minor loops. Due to the strength and direction of the effective fields, the minor loop shifts horizontally with different amplitudes of the current. The horizontal shift is recorded by the changing switching fields and is represented by the data points shown in the switching phase diagrams in Fig. 3. In Fig. 3a we show the phase diagram of a special x-type W device where the longer axis of the MTJ pillar is parallel with the channel direction, as introduced by Fukami et al.[9]. In this geometry due to the parallel direction, current induced effective fields, either field-like field or Oersted field, if present, is perpendicular to pillar's anisotropy direction, which causes them to have minimal effect on the switching behavior. The phase diagram shows a symmetric diamond shape where large current in the channel reduces the coercivity of the FL mainly through Joule heating[10]. Since Joule heating does not have a preferential polarity, the reduction of coercivity is symmetric about the center of the minor loop and is equivalent for both positive and negative current. In other words, upon applying increasing current in the SHE channel, the center of the minor loop (blue points) almost stays unchanged, which is indicative of close-to-zero effective field. Figure 3b shows a regular y-type MTJ device fabricated the same ways as in the W samples, but with a Ru channel with a stack Ti(1)-Ru(5)-FeCoB(1.4)-MgO(1.6)-FeCoB(4)-Capping. Since Ru is not known to have observable spin Hall effect (no current induced switching as shown in supplemental material), The reduction of the coercive field is mainly caused by Joule heating and Oersted field in this geometry. The Ru sample shows a phase diagram slightly deviated from the diamond shape, due to the additional effect from the Oersted field favoring a certain direction of the switching at a certain current polarity. Figure 3c shows a y-type W device with strong spin torques. With large net effective field[2], behavior at high electric current shows further deviation from diamond shape, with one side of the switching changing dramatically due to ( effective field + heating ) and the other side almost staying at the same field due to ( effective field – heating ). The comparison between the three types of devices shows clearly the effectively field in W samples, and a significant change of switching behavior when it is present.

To further explore the role of the effective field in pulse switching, we fabricate another sample Ta(1)-$Pt_{85}Hf_{15}$(4)-Pt(1)-Hf(0.3)-FeCoB(1.3-1.6)-MgO-Capping (hereafter called PtHf samples). With the same reference layer direction, the opposite directions of the current switching loops between PtHf and W samples indicate opposite spin Hall signs, consistent with previous observations[2,11,12]. We apply an external field equal to the dipole field to cancel the residual impact from the reference layer. Table 3 shows results from a series of devices on both PtHf and W samples (more statistics in supplemental materials). The “l”(“r”) in the device name indicates a field minor loop pointing to the left (right) as shown in Fig. 4a and 4b. The reversal of the minor loop direction is done through applying a large external field (> 2000 Oe) that reverses the reference layer direction. With the same device but opposite reference layer directions, the switching polarity changes but the actual directions of the FL switching are the same for a given direction of the current. For example, in a PtHf “r” device, positive current represents a P-AP switching but in a PtHf “l” device, positive current represents an AP-P switching. However, in both cases the FL switches from ← to → due to the same spin polarization determined by the current direction. Our results show that in a PtHf “r” device P-AP is favored with smaller $V_0$ and in a PtHf "l" device AP-P is favored. This has indicated that there is an extra effective field in the system that is in the direction of →. The known current induced fields reverse direction when current direction is reversed, which causes them to have the same effect (favor or hinder) on both polarities of the switching in a current switching loop. Thus the favoring of one single direction of the FL switching indicates an effective field that is in a fixed direction. The results of W devices (Table 3 and Fig. 3d) show the similar property but an opposite direction of the effective field. These seem to point to an extra effective field that is material system specific or possibly asymmetric spin torque with opposite current. While fast and reliable switching has been achieved in 3T-MTJ systems which shows great potential for cache memory application, the existence of this field confirms the sensitive nature of the SOT switching in nano-structures and more detailed modeling or testing is necessary for further utilizing 3T-MTJ structures in spin logic or in-memory computing technologies.

In conclusion, we have shown in this letter that tuning external field in pulse measurements of 3T-MTJs induces symmetry changing of the FL switching behavior in two ways. When there is a large change of the field, critical voltage changes due to the remaining effective field helping or hindering the switching. When the change of field is small, characteristic switching time changes due to the modification of initial magnetization configuration. Switching phase diagrams show clear evidence of current induced effective field in y-type 3T-MTJs with large SHE materials. We have also shown that there is an extra effective field in W and $Pt_{85}Hf_{15}$ systems with opposite directions.

This finding should confirm the complex nature of the switching dynamics and stimulate further research on the details of the switching in order to apply the SOT MTJ schemes to broader range of applications.

**Acknowledgements**

This report is based upon work supported by the Office of the Director of National Intelligence (ODNI), Intelligence Advanced Research Projects Activity (IARPA), via contract W911NF-14-C0089. The views and conclusions contained herein are those of the authors and should not be interpreted as necessarily representing the official policies or endorsements, either expressed or implied, of the ODNI, IARPA, or the U.S. Government. The U.S. Government is authorized to reproduce and distribute reprints for Governmental purposes notwithstanding any copyright annotation thereon. Additionally, this work was supported by the NSF/MRSEC program (DMR-1120296) through the Cornell Center for Materials Research, by the Office of Naval Research, and by the NSF (Grant No. ECCS-0335765) through use of the Cornell NanoScale Facility/National Nanotechnology Coordinated Infrastructure.

Figure 1

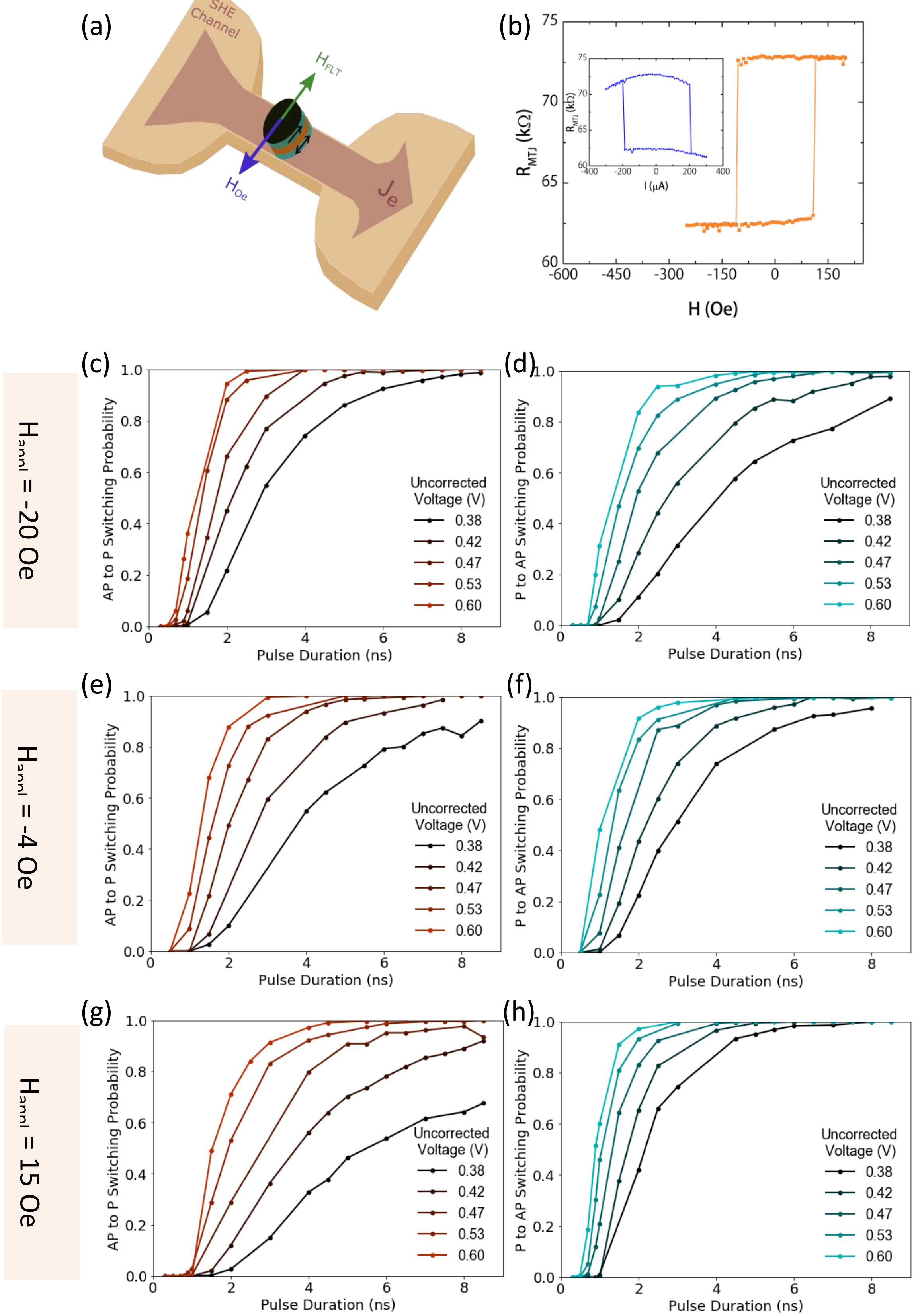

(a)
SHE Channel
H_FLT
H_Oe
J_e
(b)
R_MTJ (kΩ)
H (Oe)
-600 -450 -300 -150 0 150
60 65 70 75
R_MTJ (kΩ)
I (μA)
-400 -200 0 200 400
(c)
(d)
(e)
(f)
(g)
(h)
H_appl = -20 Oe
H_appl = -4 Oe
H_appl = 15 Oe
AP to P Switching Probability
P to AP Switching Probability
Pulse Duration (ns)
Uncorrected Voltage (V)
0.38
0.42
0.47
0.53
0.60

Figure 2

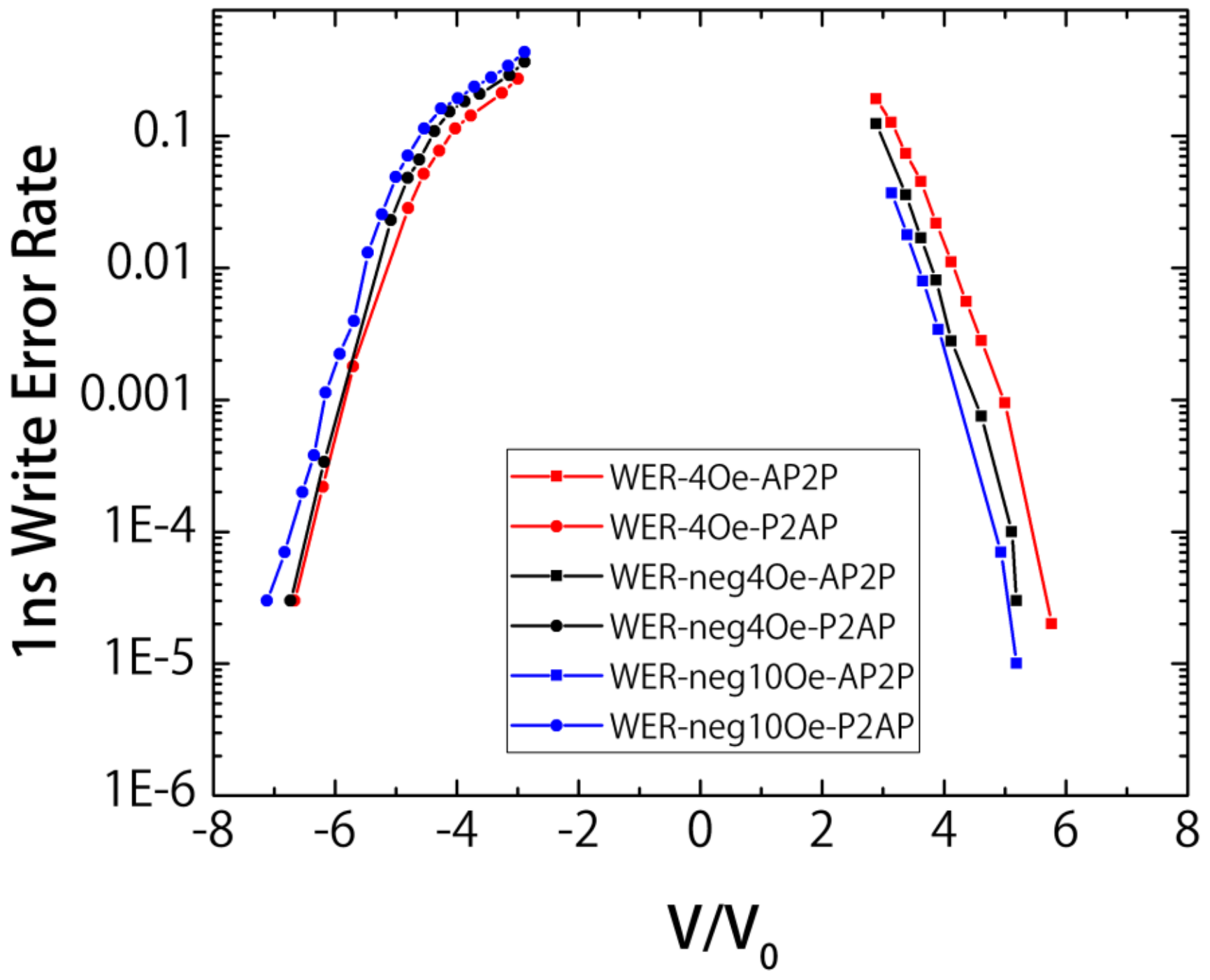

Figure 3

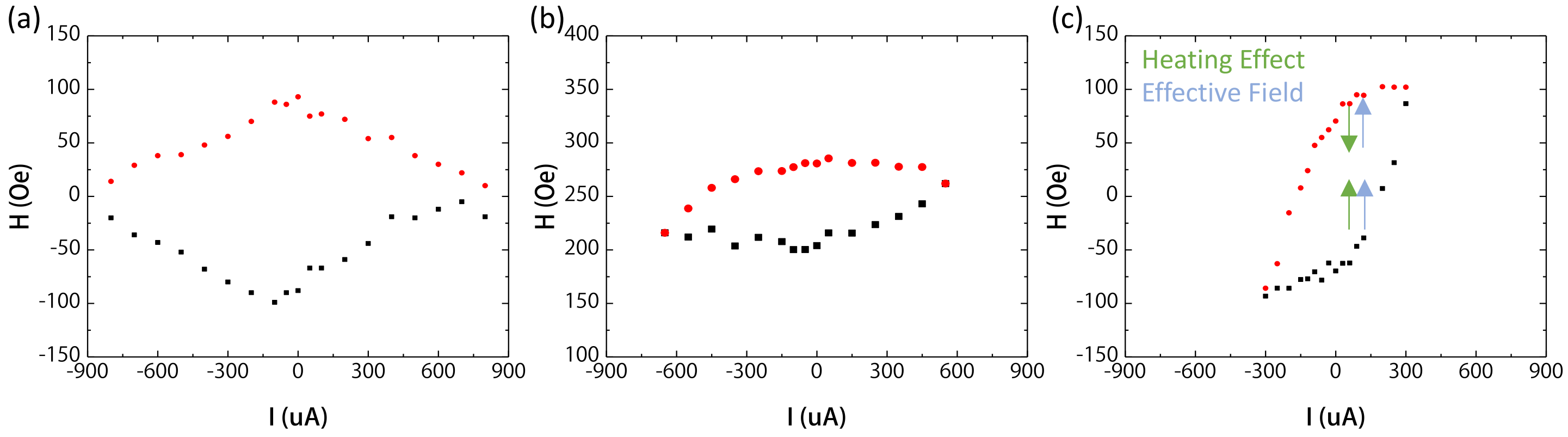

Figure 4

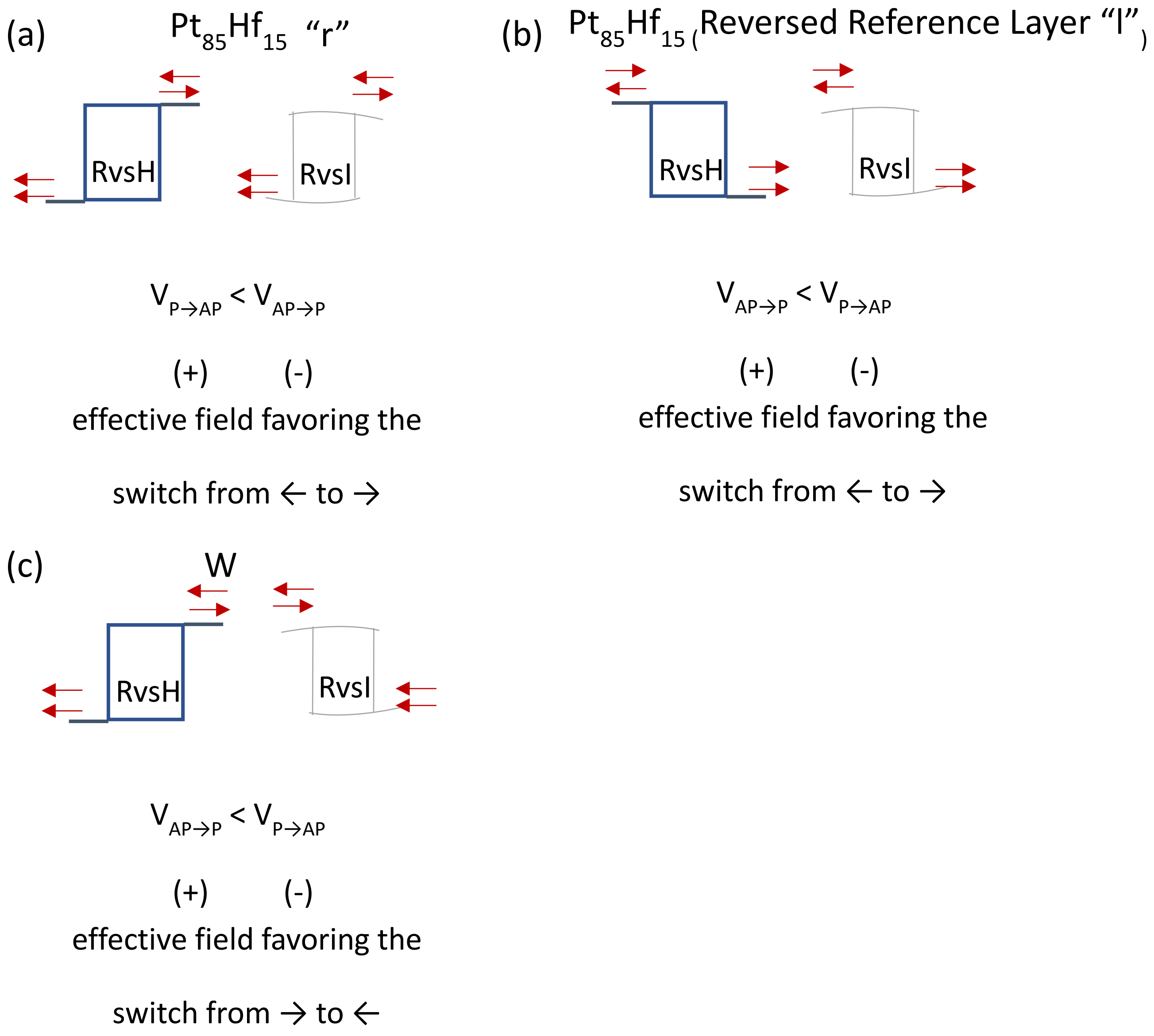

(a)
Pt85Hf15 "r"
RvsH
RvsI
VP→AP < VAP→P
(+) (-)
effective field favoring the
switch from ← to →
(b)
Pt85Hf15 (Reversed Reference Layer "l")
RvsH
RvsI
VAP→P < VP→AP
(+) (-)
effective field favoring the
switch from ← to →
(c)
W
RvsH
RvsI
VAP→P < VP→AP
(+) (-)
effective field favoring the
switch from → to ←

Table 1

Critical switching voltage and time under large change of applied external field.

| $H_{appl}$ | AP-P | | | P-AP | | |
|---|---|---|---|---|---|---|
| | $t_0$ | $V_0$ | $E \sim 4V_0^2t_0$ | $t_0$ | $V_0$ | $E \sim 4V_0^2t_0$ |
| -20 Oe | 2.17 | 0.21 | 0.383 | 1.57 | 0.27 | 0.458 |
| -4 Oe | 1.64 | 0.26 | 0.443 | 1.32 | 0.26 | 0.357 |
| 15 Oe | 1.61 | 0.29 | 0.542 | 1.34 | 0.23 | 0.284 |

Table 2

Critical switching voltage and time under small change of applied external field.

| $H_{appl}$ | AP-P | | | P-AP | | |
|---|---|---|---|---|---|---|
| | $t_0$ | $V_0$ | $E \sim 4V_0^2t_0$ | $t_0$ | $V_0$ | $E \sim 4V_0^2t_0$ |
| -10 Oe | 1.56 | 0.26 | 0.422 | 1.41 | 0.26 | 0.381 |
| -4 Oe | 1.64 | 0.26 | 0.443 | 1.32 | 0.26 | 0.357 |
| 4 Oe | 1.81 | 0.26 | 0.489 | 1.25 | 0.25 | 0.313 |

Table 3

Critical switching voltage and time under small change of applied external field.

| Device | $Pt_{85}Hf_{15}$ | | Device | W | |
|---|---|---|---|---|---|
| | $V_0$ – AP2P | $V_0$ – P2AP | | $V_0$ – AP2P | $V_0$ – P2AP |
| 1 r | -0.36 | 0.32 | 1 r | 0.22 | -0.34 |
| 2 r | -0.43 | 0.25 | 2 r | 0.23 | -0.26 |
| 3 r | -0.25 | 0.16 | 3 r | 0.26 | -0.34 |
| 4 l | 0.24 | -0.30 | 4 l | -0.36 | 0.30 |
| 5 l | 0.22 | -0.26 | 5 l | -0.34 | 0.21 |
| 6 l | 0.19 | -0.25 | 6 l | -0.30 | 0.23 |

**Figure 1 | Nanosecond pulse switching measurements with different applied external field.**

**a**, A regular y-type device with MTJ perpendicular to the underlying spin Hall channel. **b**, Magnetic minor loop of a W device. Inset: Current induced switching loop of the same device. **c and d.** Pulse switching probability curves for AP-P and P-AP under applied field of -20 Oe. **e and f**, Pulse switching probability curves for AP-P and P-AP under applied field of -4 Oe. **g and h**, Pulse switching probability curves for AP-P and P-AP under applied field of 15 Oe.

**Figure 2 | WER measurements on a W device with small change of applied external field.**

Write error rate measured at 1 ns pulse duration at three different applied fields, -10 Oe, -4 Oe and 4 Oe. A clear change in WER symmetry is seen when varying field at small steps, when $V_0$ stays almost unchanged.

**Figure 3 | Switching phase diagram in various samples.**

Switching fields under different bias current in the spin Hall channel are plotted in each phase diagram graph. **a**, W sample with device parallel to the spin Hall channel (x-type). **b**, Ru sample with device perpendicular to the spin Hall channel (y-type). **c**, W sample with device perpendicular to the spin Hall channel (y-type).

**Figure 4 | Effective field and switching schematics for $Pt_{85}Hf_{15}$ and W samples**

**a**, Field and current switching schematics for $Pt_{85}Hf_{15}$ samples with an “r” type minor loop. **b**, Field and current switching schematics for $Pt_{85}Hf_{15}$ samples with an “l” type minor loop. **c**, Switching direction for W samples with an “r” type minor loop.